\documentclass[copyright,creativecommons]{eptcs}
\providecommand{\event}{The 3rd International Workshop on \\ Trends in Functional Programming in Education, TFPIE 2014} % Name of the event you are submitting to
\usepackage{breakurl}
\usepackage{eclbkbox, proof, slashbox, ulem, soul, alltt, amssymb,
  color, booktabs, dcolumn, graphicx, setspace}             % Not needed if you use pdflatex only.

\newcommand\AND{\texttt{\symbol{"26}}} % &
\newcommand\UP {\texttt{\symbol{"5E}}} % ^
\newcommand\US {\texttt{\symbol{"5F}}} % _
\newcolumntype{d}[1]{D{.}{.}{#1}}

\title{Report on a User Test and Extension of a Type Debugger for Novice Programmers}
\author{Yuki Ishii
\institute{Department of Information Science,\\  Ochanomizu University\\ Tokyo, Japan}
\email{ishii.yuki@is.ocha.ac.jp}
\and
Kenichi Asai
\institute{Department of Information Science, \\ Ochanomizu Univeristy\\
Tokyo, Japan}
\email{\quad asai@is.ocha.ac.jp}
}
\def\titlerunning{Report on a User Test and Extension of a Type Debugger for Novice Programmers}
\def\authorrunning{Yuki Ishii \& Kenichi Asai}
\begin{document}
\maketitle

\begin{abstract}
A type debugger interactively detects the expressions that cause type
errors.
It asks users whether they intend the types of identifiers to be those
that the compiler inferred.
However, it seems that novice programmers often get in trouble when
they think about how to fix type errors by reading the messages given
by the type debugger.
In this paper, we analyze the user tests of a type debugger and
report problems of the current type debugger.
We then extend the type debugger to address these problems.
Specifically, we introduce expression-specific error messages and
language levels.
Finally, we show type errors that we think are difficult to explain to
novice programmers.
The subjects of the user tests were 40 novice students belonging to the
department of information science at Ochanomizu University.
\end{abstract}

\section{Introduction}
Strongly-typed languages, such as OCaml or Haskell, provide
programmers with type safety via static type checking at compile time.
However, it is not always easy for programmers, especially novices, to write well-typed programs.
In particular, the error messages provided by the compiler often do
not indicate the source of the type error.
For example, the OCaml compiler prints the following message, when a
programmer tries to define a function that calculates the \texttt{x}-th
power of \texttt{(x + 1)}: 
\begin{spacing}{1.0}
\begin{alltt}
fun x -> \underline{(x + 1)} ^ x
Error: This expression has type int
       but an expression was expected of type string
\end{alltt}
\end{spacing}
This error message says that the type of \texttt{x + 1}
conflicts with the type of the first argument of \UP ,  which concatenates two string values in OCaml.
If the programmer blindly follows the error message and changes the
type of \texttt{x + 1} to \texttt{string} (e.g., by inserting
\texttt{string\US of\US int}), he or she will end up with a different program
than originally intended:
\begin{verbatim}
fun x -> (string_of_int x + 1) ^ (string_of_int x)
\end{verbatim}

Since the compiler does not know the intention of the programmer, it
is unable to show a single error message
  that reflects that intention.
The above error message simply reports the fact that the two types are
in conflict during type inference.
To remedy this situation, Chitil \cite{Chitil2001} proposed an interactive type
debugger.
Using algorithmic program debugging \cite{Shapiro1983} on the compositional type inference
tree, this sort of debugger ascertains the programmer's intention by
asking questions and detecting the sources of the type errors.
Tsushima and Asai \cite{Tsushima2012IFL} followed up on this work by implementing a
type debugger for full OCaml by reusing its type inferencer.

In 2012, we used the OCaml type debugger in a ``Functional
Programming'' course at Ochanomizu University and collected logs
showing how
students interacted with the type debugger.
In this paper, we report on the results of analyzing the logs, describe
problems of the type debugger, and extend the
type debugger accordingly to make it novice-friendly.

This paper is structured as follows.
In Section 2, we review how our type debugger works.
In Section 3, we analyze the logs of the type debugger and discuss the
results in Section 4.
We extend the type debugger in Section 5 and evaluate it in Section
6. Its limitations are discussed in Section 7.
Related work is in Section 8, and we conclude
the paper by outlining future work in Section 9.

\section{Type debugger}

Let us review how the type debugger works.
The type debugger constructs the most general type tree
\cite{Chitil2001,Tsushima2012IFL}
and uses algorithmic program debugging \cite{Shapiro1983} to detect type errors.

\subsection{Most General Type Tree}

The type debugger uses the most general type tree (MGTT) to detect the
source of a type error. Unlike the standard type inference tree used
in a compiler, an MGTT maintains the most general type for each
subexpression. For example, the MGTT for the previous program becomes
as follows:\  (Here, we have abbreviated the types of \texttt{+} and \texttt{\^} as {\tt
  $\tau^+$ $=$ int $\rightarrow$ int $\rightarrow$ int} and {\tt
  $\tau^{\wedge}$ $=$ string $\rightarrow$ string $\rightarrow$
  string}, respectively.) \\
%The most general type tree (MGTT) which the type debugger uses is the compositional type tree \cite{Chitil2001} devised by Chitil. The MGTT of the previous program is as shown below (where {\tt $\tau^+$ $=$ int $\rightarrow$ int $\rightarrow$ int} and {\tt $\tau^{\wedge}$ $=$ string $\rightarrow$ string $\rightarrow$ string}). \\
\begin{center}
\small{
\infer{\texttt{fun x $\rightarrow$ (x + 1) \^\ x $\cdots$ type error }}
      {\infer{\texttt{\{\} $\vdash$ (x + 1) \^\ x $\cdots$ type error }}
             {\infer{\texttt{\fbox{\{x :\ int\}}$^B$ $\vdash$ (x + 1) :\ int}}
                    {\texttt{\fbox{\{x :\ a\}}$^A$ $\vdash$ x :\ a}
                            & {\texttt{\fbox{\{\}}$^A$ $\vdash$ + :\ $\tau^+$}}
                            & {\texttt{\fbox{\{\}}$^A$ $\vdash$ 1 :\ int}}}
                          & {\texttt{\{\} $\vdash$ \^\ :\ $\tau^{\wedge}$ }}
                          & {\texttt{\{x :\ b\} $\vdash$ x :\ b }}}}
}
\end{center}
The MGTT is different from the standard type inference tree in that
the information that \texttt{x} has type \texttt{int} in box $B$
does not propagate to boxes $A$.
To type the three subexpressions, \texttt{x}, \texttt{+}, and
\texttt{1}, independently, we do not need to constrain the type of
\texttt{x}.
The type of \texttt{x} becomes \texttt{int} only when these
subexpressions are composed and \texttt{x} is passed as an argument to
\texttt{+}. This kind of bottom-up type inference was used in ${\cal U}_{AE}$
  \cite{Yang2000} as well as in the Helium compiler \cite{Hage2003} to remove
  the left-to-right bias of the type inference and produce better error
  messages.

MGTT is compositional: the most general type of an expression is
determined solely from the expression and does not depend on other
expressions.
By comparing the most general type with the programmer's intended type, we can
detect the source of a type error \cite{Chitil2001}.
In this paper, we say that an expression has a \emph{well-intended}
  type if the type of the expression does not contradict the
  programmer's intention.

For example, suppose that a programmer intends the above function to have one
of two types:
\begin{enumerate}
\item {\tt int -> string}
\item {\tt int -> int}
\end{enumerate}
In the first case, the programmer's intended program is
\texttt{fun x -> string\US of\US int (x + 1) \^{} \\ string\US of\US int x}.
While the type of \texttt{x} in boxes $A$ does not contradict the
programmer's intention, the type of \texttt{x} in box $B$ does
contradict them.
In other words, \texttt{x}, \texttt{+}, and \texttt{1} all have the
  well-intended types, but \texttt{(x + 1)} does not. Thus, we conclude that \texttt{x + 1} is the source of the type
error. 

If we used the standard type inference tree instead, we could detect
that the conflict first occurs somewhere in boxes $A$ or $B$.
However, we cannot further identify the source of the type error,
because the type of \texttt{x} propagates to the boxes $A$ via
unification and we have no information when the type of {\tt x} is
first forced to {\tt int}.

On the other hand, suppose that the programmer intends the second
type. In this case, the intended program is {\tt fun x -> power (x + 1) x}
where {\tt power x y} calculates the x-th power of {\tt y}. Since the
programmer intends {\tt \^{}} to be {\tt int -> int -> int}, the
actual type of {\tt \^{}} conflicts with the programmer's intention. Therefore, we can detect that {\tt \^{}} is the
source of the type error.

\subsection{Algorithmic Programming Debugging}

The type debugger detects the source of a type error by using
algorithmic program debugging (APD) to traverse the MGTT \cite{Shapiro1983}.
APD was originally devised by Shapiro to find an error in a
Prolog program. It can be used to detect errors in any tree structure.
The algorithm starts from a node with an error and proceeds as follows.
\begin{enumerate}
\item Check whether any of its child nodes has an error.
\item If no child node has an error, the current node is the
source of the error.
\item If one of the child nodes has an error, apply APD to the child
node. 
\end{enumerate}
In the last step, if two or more child nodes have an error, one
  of the erroneous child nodes is chosen.
The final result depends on which one is chosen.

\subsection{Detecting type errors}
Our type debugger detects the source of a type error in two steps:
\begin{enumerate}
\item Find the node (expression) that has a type error, but all of its
child nodes are well-typed.
\item Find the node (expression) that does not have well-intended
  types, but all of its child nodes have well-intended types.
\end{enumerate}
In the first step, the type debugger uses the compiler's type
inferencer to judge whether a node has a type error.
In the second step, the type debugger asks the programmer whether the
types inferred by the compiler's type inferencer match his/her
intention.
In particular, it asks whether the environments and expressions are of
the intended types.

In the MGTT in Section 2.1, for example, the type debugger starts from
the bottom of the tree and reaches node \texttt{(x + 1) \UP\ x}
as a result of executing step 1, because all of its child nodes are well
typed.
Starting from this node, the type debugger asks the programmer if each
subexpression has the intended type.

If the programmer answers that all the child nodes
have intended types, the current expression is identified as the source of the
type error, because the type of \texttt{x + 1} does
not match the type of the first argument of \UP.
In this case, the identified expression is ill-typed.

On the other hand, if the programmer answers that the type of
\texttt{x} should be {\tt string} (in box $B$), the node \texttt{x + 1} is identified as
the source of the type error, because the type of \texttt{x} is first
forced to \texttt{int} here.
In this case, the identified expression is well-typed, but not well
intended.

\section{Analysis}

In the spring of 2012, we used the type debugger in a ``Functional
Programming'' course offered at our university.
The course was taken by 40 CS-major students.
Although they had one-year of experience writing C programs, it was
the first time for them to write programs in a strongly-typed language.
During the course, the students learned OCaml and wrote a solution to
the shortest path
problem for the Tokyo metro network.
We instructed students to use the type debugger when they encountered
type errors.
When (and only when) they used the type debugger, we collected the erroneous programs and
their interactions with the type debugger. 

We analyzed the type-error logs in two ways:\
\begin{enumerate}
\item Which expression was detected as the source of the type error?
\item How did the students change their programs after reading the error message? 
\end{enumerate} 

\subsection{Expressions identified as sources of type errors}

Table \ref{tab:error1} shows a breakdown of expressions identified as sources of
type errors by the type debugger. We collected 704 logs and classified them
manually. Among the seven categories, the identified expression was ill-typed in
the first five categories and well-typed in the last two categories.

In this section, we describe typical type errors from the logs and
analyze them to see if the type debugger worked effectively.

\begin{table}[t]
  \begin{center}
    \begin{tabular}{|l|d{2.0}|} \hline
      expression & \%  \\ \hline \hline
      Application & 30.2  \\ \cline{1-2}
      Match expression & 13.5  \\ \cline{1-2}
      Constructor & 11.4  \\ \cline{1-2}
      Conditional & 4.3  \\ \cline{1-2}
      Recursive function & 3.7  \\ \hline
      Environment & 18.7  \\ \cline{1-2}
      Syntax misunderstood & 18.2  \\ \hline
    \end{tabular}
    \caption{Classification by expression}
    \label{tab:error1}
  \end{center}
\end{table}

\paragraph{Application.}

30.2\% of the sources of type errors were located in the application.
After an application expression was identified, the type debugger determined
which of the arguments caused the type error.
It passed an increasing number of arguments, starting from the first
one, to the function.
The first argument that caused a type error was shown to the programmer
as the conflicting argument.

For example, a typical type error is shown below.
The box in the program shows the highlighted part. 
\begin{spacing}{1.0}
\begin{alltt}
fun x -> \fbox{(x + 1) ^ x}\ensuremath{\sp{1}}
Error:
The first argument of this application causes a type error. {\rm (highlight 1)}
\end{alltt}
\end{spacing}
In the error message, ``\texttt{this application}'' refers to the
application of the {\tt \^{}} operator and ``\texttt{the first
argument}'' refers to \texttt{x + 1}.
However, since the {\tt \^{}} operator uses infix notation, most students
had trouble understanding which expression ``\texttt{this application}''
refers to. 

Moreover, even after students understood what ``{\tt this application}'' refers to,
the error message was still not very helpful: 
\begin{alltt}
(* f :\ int list -> int -> int list *)
(* g :\ int list -> int list *)
let test = \fbox{f (g lst) = [a; very; large; list;
...]}\ensuremath{\sp{1}}
Error:
The second argument of this application causes a type error. {\rm (highlight 1)}
\end{alltt}
Although this error message suggests that ``\texttt{[a;\ very;\
large;\ list;\ ...]}'' is the source of the type error, it is actually
not.
What caused this type error is that the student passed only one
argument to function \texttt{f}, which required two.
It resulted in the type of \texttt{=} being
\texttt{(int -> int list) -> (int -> int list) -> bool}.

Since the error message mentions only the second argument, the students
tended to check only the large list and rarely found they had forgotten an argument
for \texttt{f (g lst)}.
Even if the student could understand the error message, it seemed to
be of little help in finding and fixing the source of the type error.

\paragraph{Match expression.}
As the course progressed, the programs that the students wrote became
more complicated.
One sort of the complex expressions that the students wrote is match expressions.
Here is an example where a student struggled to correct a type error
more than ten times.\\
\\
\texttt{type station\US t = \{start :\ string; destination :\ string; distance :\ float;\}} \\
\texttt{type tree\US t = Empty | Node of tree\US t * string * (string * float) list * tree\US t} \\
\texttt{let rec insert\US station station\US tree station = }
\begin{breakbox}
\noindent \hspace{4mm} \texttt{match station with} \hspace{118mm} $^3$
\par \texttt{[] -> \fbox{[]}$^1$ }
\par \texttt{|\{start = st; destination = dest; distance = dist;\} ::\ rest -> } 
\begin{breakbox}
\par \quad \texttt{match station\US tree with } \hspace{98mm} $^2$
\par \quad \quad \texttt{Empty -> Node (Empty, st, [(dest, dist)],Empty) }
\par \quad \quad \texttt{| Node (t1, name, station\US list, t2) -> }
\par \quad \quad \quad \quad \texttt{if name < st then }
\par \quad \quad \quad \quad \quad
  \texttt{Node (t1, name, station\US list, insert\US station t2 station) }
\par \quad \quad \quad \quad 
  \texttt{else if name > st then }
\par \quad \quad \quad \quad \quad
  \texttt{Node (insert\US station t1 station, name, station\US list, t2) }
\par \quad \quad \quad \quad \quad \quad
  \texttt{else insert\US name station\US tree rest }
\end{breakbox}
\end{breakbox}
{\tt \noindent Error:\ \\
 This match expression causes a type error.} (highlight 3) \\
\\
The type error occurred because the match expression (in highlight
3) returns an empty list in highlight 1 but a tree in highlight 2.
However, after asking various questions, the type debugger identified
the whole match expression (highlight 3) as the source of the type
error and responded with an uninformative error message.
Because a large area (highlight 3) was identified as the source of the
type error, the student had a hard time finding out which part of the
highlight was wrong.

The source of the type error was identified as the whole match
expression, because the type debugger works at the level of
expressions.
It first checks whether all the subexpressions are well-typed: i.e.,
it checks whether the expression to be matched (\texttt{station})
as well as all the branches of the match expression are well-typed.
When the student answers that their types are as intended,
the debugger concludes that type constraints of this match
expression are not satisfied.
Working at the level of expressions has the benefit that we can build a
type debugger without adding any expression-specific type constraints.
If we had to consider all the type constraints for all the
expressions, it would have been difficult to support all the
constructs in OCaml.
However, the fact that the students had to struggle debugging the above
program shows that we should actually consider expression-specific
type constraints
for common cases so that we can provide more informative error messages.

\paragraph{Conditional expression.}
The same problem applies to conditional expressions.
Whenever a conditional expression is identified as the source of a
type error, the type debugger shows the same error message.
However, a conditional expression has its own type constraints:
the predicate part has to have type \texttt{bool} and the two branches
must have the same type.
In these cases, it would be more informative to show them in the error
message.

Another common mistake regarding conditional expressions happens when
students forget to write the \texttt{else} branch.
In this case, the \texttt{then} branch must have the \texttt{unit} type.
Even if students could answer the questions asked by the type debugger
correctly, they might still be puzzled by the error message,
saying that the \texttt{then} branch must have the \texttt{unit} type,
because the \texttt{unit} type is introduced in a later
stage of the course; they would not know what a \texttt{unit} is at this
point.

\paragraph{Constructor.}
When students define a new data type, the types of the arguments are often
wrong.
For example, suppose we have a tree with \texttt{char} and
\texttt{int} information at each node:
\begin{verbatim}
type tree_t = Empty
            | Node of tree_t * char * int * tree_t
\end{verbatim}
If a student forgets to write one of its argument, e.g.,
\texttt{Node (left, n, right)}, the type debugger identifies
this expression to be the source of the type error: after confirming
that the student intended the
tuple to have the type \texttt{tree\US t * int * tree\US t}, it reports
that the number of arguments of \texttt{Node} is four.
However, since the definition of \texttt{Node} is usually placed far
before its use, the students would not immediately see why four
arguments instead of three are required.

Because constructors are not functions and are handled differently
from functions in OCaml, the current implementation of the type
debugger does not ask the intended type of constructors, but rather assumes
that the types are as intended.
The above problem could have been avoided if the type debugger asked the intended type of
the constructor and jumped back to its definition when it
was different from the inferred type.

\paragraph{Recursive functions.}
In some cases, students try to use a recursive function of a different type
from its definition.
In such case, the current error message shows the two types of
recursive function:
the type inferred from the outermost structure
and the type of its use.
For example, for the following function where the second argument to
\texttt{gcd} is missing at the recursive call:
\begin{alltt}
let rec gcd \fbox{m n = if n = 0 then m else gcd n} \ensuremath{\sp{1}}
\end{alltt}
the type debugger shows the following error message:
\begin{alltt}
While this expression has 'a -> int -> 'a,
you tried to use it as type int -> 'a for recursive call. {\rm (highlight 1)}
\end{alltt}
\paragraph{Well-typed but unintended expressions.}
If a student answered that the type shown in the MGTT is different
from his or her intention, the type debugger traverses the MGTT to the
node whose expression is well-typed.
This case can be divided into two subcases.

One subcase is when the type of a variable in the environment is not the
intended one.
This situation is listed in Table 1 as ``Environment''.
Here, the first node where the type of the variable is forced
as such is identified as the source of the type error.
From that, students can understand why the variable's type is not as
intended.

The other subcase is when students misunderstand (or mistakenly use)
syntax.
This situation is listed in Table 1 as ``Syntax misunderstood''.
For example, if a student writes a floating point number $1.2$ as
\texttt{1,2}, it is well-typed as \texttt{int * int}, but not as
\texttt{float}.
Another example is to write a list of numbers \texttt{[1;\ 2]} using a
comma \texttt{[1, 2]}. The latter is a singleton list of a pair of integers.
In these cases, the type debugger shows the type of the expression,
which is enough for students to understand the cause of the type error
most of the time.

\begin{table}[t]
  \begin{center}
    \begin{tabular}{|l|c|c|c|c|} \hline
      & Effective & \shortstack{\\[1mm]appropriate expression,\\but ineffective} & unrelated expression & No changes \\ \hline \hline
      Application & 36.3 & 18.2 & 12.5 & 33.0 \\ \hline
      Match expression & 38.0 & 10.3 & 17.2 & 34.5 \\ \hline
      Constructor & 20.7 & 17.2 & \ 6.9 & 55.2 \\ \hline
      Conditional & 0 & 50.0 & 0 & 50.0 \\ \hline
      Recursive function & \ 7.7 & \ 7.7 & 23.1 & 61.5 \\ \hline
    \end{tabular}
    \caption{Analysis by reaction of students (\%)}
    \label{tab:error2}
  \end{center}
\end{table}

\subsection{Students' reactions}
For the first five cases of Table 1, we classified how the students
changed their program after they had read an error
message from the type debugger in four ways,
following Marceau et al.\ \cite{Marceau2011.10}:

\begin{enumerate}
\item An effective change that fixes the type error and brings the
  student closer to a solution.
\item An ineffective change to the appropriate expression.
\item A change to an expression that is unrelated to the type error.
\item No changes, such as inserting indentation or spaces.
\end{enumerate}
We can judge the effectiveness of changes, because we know what the
student is aiming for: the solutions to the assignments.
In addition, we use the purpose statement the student writes to guess
his intention.
In the course, students were advised to follow the design recipe
\cite{HtDP} and write a purpose statement together with its type for
each function definition.
When we classified the students' reactions, we used this information to
judge if the change was effective.
However, the type they write in the purpose statement is often wrong.

Table 2 shows the classification of the students' reactions.
We found that not many programs could be corrected effectively.
Although students could fix nearly 40\% of the type errors in
applications or match expressions, they could not make any
changes to a third of them. Furthermore, more than 60\% of the type errors
located in recursive functions were left unchanged.

We analyzed the application, match expressions, and conditional
expressions as follows.

\paragraph{Application.}

When an application is detected as the source of a type error,
the type debugger prints the argument that caused the type error.
As a result, the students often changed that argument or swapped the
arguments.
However, they seemed to get in trouble when the infix notation was
used,
or the argument was unintentionally higher-order (such as we saw in Section 3.1).

\paragraph{Match expression.}

Students tended to change the wrong expression when the highlight
covered a
wide area of the program. Since match expressions often include
complicated data types, students sometimes changed those complicated
expressions to simpler ones without type errors and the students could
identify where to fix them. 
This accounted for 38\%, a relatively high percentage, of the effective changes.

\paragraph{Conditional expression.}

As we saw in Section 3.1, students tended to get in trouble when they
forgot to write {\tt else} statements. In this case, they
often changed the {\tt then} statement into something else and rarely
noticed that what they needed to do was add an {\tt else} statement.

\section{Discussion}

The analysis in Section 3 shows that there is plenty of room for
improvement when it comes to helping novice programmers use the type
debugger effectively.
Our main goal here is to help students understand
type errors and learn how to program in typed languages in general, by
reading error messages of the type debugger.
In this section, we discuss the problems of the current type debugger
as regards this goal.

The first problem is that the type debugger does not
show enough information for programmers to understand why the type errors
occurred.
As the second program of application in Section 3.1
shows, students could not find the actual source of the type error
in the first argument, because the type debugger said the second
argument caused the type error.
Thus, rather than simply showing the first type-conflicting argument, we
should show more information, such as the types of other arguments, so that
students can consider what went wrong.

The second problem is that the detected expressions are sometimes too
large for students to find the source of the type error.
If the detected expression is ill-typed and violates a type
constraint, the type debugger should show it so that students can see
immediately what constraint is violated.

In light of these results and considerations, we decided to extend the type debugger in three ways.
\begin{enumerate}
\item Smaller highlights: \\
   When ill-typed conditional or match expressions are detected as the
   source of the type error, we use expression-specific type
   constraints to narrow the highlighted part.  This also leads to
   novice-friendly error messages.

\item Novice-friendly error messages: \\
   In addition to expression-specific type error messages, we provide
   detailed information why the type error occurred, in particular for
   applications.

\item Introduction of language levels \cite{Felleisen2004} (as in Racket\footnote[1]{http://racket-lang.org/}): \\
   To provide appropriate error messages, we prohibit the use
   of advanced language features that students are not supposed to
   use.
\end{enumerate}
We describe these extensions in the next section.

\section{Extensions}

\subsection{Expression-specific error messages}

As we saw in Section 2, the first step of type debugging detects a
node in a tree that is ill-typed but whose child nodes are all well
typed.
If types of all its subexpressions are as intended, the type
debugger reports the current expression as the source of the type
error.
When the detected expression itself is ill-typed, however, we want to
have a detailed explanation as to why it is ill-typed.

To provide programmers with more informative error messages, we
introduce expression-specific error messages for conditional
expressions, match expressions, and applications.
The basic idea is to add annotations to the subexpressions of the
ill-typed expression and see if they satisfy the type constraints
imposed by the expression.

\subsubsection{Conditional expressions}

The type constraints for conditional expressions are:
\begin{enumerate}
\item The predicate part must be of type \texttt{bool}.
\item When the \texttt{else} branch is missing, the \texttt{then} branch
must be of type \texttt{unit}.
\item When the \texttt{else} branch is present, the \texttt{then} and
\texttt{else} branches must have the same type.
\end{enumerate}
To examine if these constraints are satisfied, the type debugger is
extended to perform the following checks:
\begin{enumerate}
\item The type debugger extracts the predicate part of the conditional
expression and annotates it as \texttt{bool}.  It then
passes the annotated predicate to the compiler's type inferencer.
If it reports a type error, the type debugger identifies the predicate
part as the source of the type error.
\item When the \texttt{else} branch is missing, the type debugger
extracts the \texttt{then} branch and annotates it as \texttt{unit}.
It then checks if the annotated branch type checks as in 1.
If it does not, the type debugger identifies the \texttt{then} branch
as the source of the type error.
\item When the \texttt{else} branch is present, the type debugger
constructs a list of the \texttt{then} and \texttt{else} branches and
checks if it type checks.
If it does not, the type debugger identifies that the type error
occurred because the two branches had different types.
\end{enumerate}
With this extension, the type debugger can provide better error
messages for conditional expressions. For example, before the extension, in all the following cases taken
from the logs, the type debugger reported the same error message,
saying only that the conditional expression was the source of the type
error. The newly generated messages look as follows.
\begin{enumerate}
\item Predicate:
\begin{alltt}
... if \fbox{(try kekka\US kyori with Not\US found -> infinity)} \ensuremath{\sp{1}} then ...
Error:
The type of predicate statement is float, but it should be bool. {\rm (highlight 1)}
\end{alltt}
\item Without {\tt else} statement:
\begin{alltt}
... if (a + 1) < c then \fbox{a + 1} \ensuremath{\sp{1}}
Error:
The type of then statement is int but it should be unit. {\rm(highlight 1)}
\end{alltt}
\item With {\tt else} statement:
\begin{alltt}
... \fbox{if (a + 1) < c then a + 1 else print{\US}int c} \ensuremath{\sp{1}}
Error:
The type of then statement is int and else statement is unit,
but these should be the same type. {\rm (highlight 1)}
\end{alltt}
\end{enumerate}
In the error messages, the type debugger prints the types of related subprograms and how the programmers need to change it. Also, in the first two examples, the debugger uses smaller highlights, where only detected subprograms are colored.

\subsubsection{Match expressions}

The type constraints for match expressions are:
\begin{enumerate}
\item The expression to be matched and all the patterns must have the
same type.
\item All the branches must have the same type.
\end{enumerate}
We extended the type debugger to examine these constraints, as well as the conditional expressions.
\begin{enumerate}
\item The type debugger replaces all the branches with a dummy
value (such as \texttt{()}) and checks if the resulting match
expression is well-typed.
If it is not well-typed, the type debugger identifies the first
pattern that causes a type error by repeatedly removing the last
pattern-expression pair from the list.
If the first pattern already causes a type error, it means that the
expression to be matched and the first pattern do not have the same
type.
\item If all the patterns have the same type as the expression to be
matched, the type debugger performs the same things for the original
match expression: it repeatedly removes the last pattern-expression pair
from the list to identify the first expression that causes a type error.
\end{enumerate}
We used an incremental algorithm for match expressions rather than
putting all the branches into a list (as we did for conditional
expressions), because we want to check not only whether there is a type
conflict but also which one is in conflict.
The above algorithm identifies the first conflicting branch in the
given match expression.

Using the extended type debugger, the error message of the example in
Section 3.1 (match expression) becomes as follows:
\begin{spacing}{1.0}
\begin{alltt}
Error:
The highlighted expression has type tree{\US}t and
the previous expression has type 'a list,
but these should be the same type. {\rm (highlight 2)}
\end{alltt}
\end{spacing}
In addition to printing the type of the detected subprogram in the error message, the type debugger also prints other types (such as {\tt 'a list} in the example) to help programmers easily decide which one to fix.

\subsubsection{Applications}

Since the type debugger already shows which argument causes the type
error in the error message, we did not change its basic
behavior. Instead, we changed the error message to print the types of the function,
all the arguments, and the required type for the detected expression. 

For example, for the program in Section 1 where the
programmer intends its type to be \texttt{int -> string}, we have the
following error message:
\begin{spacing}{1.0}
\begin{alltt}
fun x -> \fbox{(x + 1) \^\ x}\ensuremath{\sp{1}}
Error:
The first argument of this application causes a type error. {\rm (highlight 1)}
The types of the function, its arguments, and the required type for the first
argument are:
Function (\^{}): string -> string -> string
First argument: int
Second argument: 'a
Required for the first argument: string
\end{alltt}
\end{spacing}
\noindent The type of the second argument is \texttt{'a}, because our type
  debugger is compositional: the second argument \texttt{x} does not
  impose any constraints on the type of \texttt{x}.

\subsection{Introduction of language levels}
\begin{table}[t]
  \begin{center}
    \begin{tabular}{|c|c|c|c|c|}\hline
      level & 1 & 2 & 3 & 4 \\ \hline \hline
      \shortstack{{} \\ partial application \\ {}} & & \checkmark & \checkmark & \checkmark \\
      \hline
      \shortstack{{} \\ {\tt if} without {\tt else}, {\tt unit}, side
        effects \\ {}} & & & \checkmark & \checkmark \\ \hline
      \shortstack{{} \\ Confusing operators (\texttt{==}, \texttt{!=},
        \texttt{or}, \texttt{\&}) \\ {}} & & & & \checkmark \\ \hline
    \end{tabular}
    \caption{Language levels}
    \label{tab:lang_level}
  \end{center}
\end{table}
One of the observations gleaned from the logs is that many students forgot to
write some of the necessary arguments.
Since OCaml is a higher-order language, passing fewer arguments does
not necessarily mean a type error.
However, it often results in a complicated error message later, which
mentions higher-order types.
To provide informative error messages in such cases, we extended the
type debugger so that it would have language levels \cite{Felleisen2004}, following Racket
(See Table \ref{tab:lang_level}).

\paragraph{Level 1.}
The Level 1 language is for beginners who do not know first-class
functions.
At this level, whenever an application is identified as the source of a
type error, the type debugger checks the type of its arguments.
If any of the arguments have higher-order types, it prints out the
higher-order argument with its type and suggests that some arguments
might be missing.
This way, the type debugger can point out unintentional first-class
function values.
Such an error message can only be provided, because we assume that
students do not use partial applications at this stage.

At language level 1, the error message for the example from
Section 3.1 (application) becomes as follows: 
\begin{alltt}
(* f : int list -> int -> int list *)
(* g : int list -> int list *)
let test = \fbox{f (g lst) = [a; very; large; list; ...]} \ensuremath{\sp{1}}
Error:
The second argument of this application causes a type error. {\rm (highlight 1)}
Function (=): 'a -> 'a -> bool
First argument: int -> int list
Second argument: int list
The following arguments have the function type.
First argument: int -> int list
(some argument might be missing.)
\end{alltt}
At language levels higher than 1, the type debugger prints the
types of the function and arguments.
At language level 1, the type debugger additionally prints the types
of the higher-order arguments.

\paragraph{Level 2.}
Level 2 language is for students who know first-class functions
but do not use side effects.
Side effects are introduced only near the end of the course.
Until then, students are not allowed to use side effects and do not
know the \texttt{unit} type.

We have modified the OCaml parser to prohibit expressions related to
side effects and the \texttt{unit} type:
\texttt{if} without \texttt{else},
\texttt{for},
\texttt{while},
assignments,
and sequential execution.
At language level (1 and) 2, we can explicitly point out that
\texttt{else} is missing and suggest to add it: 
\begin{alltt}
\fbox{fun x -> if true then x + 1} \ensuremath{\sp{1}}
Error:
The else statement is missing. {\rm (highlight 1) }
\end{alltt}
\paragraph{Levels 3 and 4.}
We created one more level (level 3) before the full OCaml language
(level 4).
Language level 3 prohibits the use of confusing operators,
\texttt{==} and \texttt{!=}.
These pointer equality operators are not used in the course and
are simply prohibited.
Students are directed to use \texttt{=} and \texttt{<>}, respectively.
Students are also directed to use \texttt{||} and \AND\AND, instead of
deprecated operators, \texttt{or} and \AND, respectively.

\begin{table}[t]
\begin{center}
\begin{tabular}{@{}ld{3.0}ld{3.0}ld{3.0}@{}}
\toprule
\multicolumn{2}{c}{Conditional} & \multicolumn{2}{c}{Match} & \multicolumn{2}{c}{Application} \\
\cmidrule(l{0.75em}r{.75em}){1-2}\cmidrule(l{.75em}r{.75em}){3-4}\cmidrule(l{.75em}r{.75em}){5-6} 
Subprograms & \% & Subprograms &  \% & Subprograms & \% \\
\midrule
Predicate & 3.1 & {\tt pattern} & 0 & Argument & 98.4 \\
{\tt then} not {\tt unit} & 78.1 & {\tt exp} and {\tt pattern} & 4.9 & Difficult to explain & 1.6 \\
{\tt then} and {\tt else} & 9.4 & {\tt expression} & 80.4 &  &  \\
Difficult to explain & 9.4 & Difficult to explain & 14.7 &  &  \\
\bottomrule
\end{tabular}
\caption{Simulation of 2012 course}
\label{tab:simulation}
\end{center}
\end{table}

\begin{table}[t]
  \begin{center}
    \scalebox{0.9}{
      \begin{tabular}{@{}lcd{3.0}lcd{3.0}lcd{3.0}@{}}
        \toprule
        \multicolumn{3}{c}{Conditional} & \multicolumn{3}{c}{Match} & \multicolumn{3}{c}{Application} \\
        \cmidrule(l{0.75em}r{.75em}){1-3}\cmidrule(l{.75em}r{.75em}){4-6}\cmidrule(l{.75em}r{.75em}){7-9} 
        Subprograms & Level & \%  & Subprograms & Level &  \% & Subprograms &
        Level & \%  \\
        \midrule
        No {\tt else} & 1-2 & 1.5 & {\tt pattern} & 1-3 & 4.9 &
        Partial application & 1-2 & 6.6 \\
        Predicate & 1-3 & 0 &{\tt exp} and {\tt pattern}  & 1-3 & 3.2 & Argument & 1-3 & 75.6 \\
        {\tt then} not {\tt unit}  & 3 & 4.5 & {\tt expression}  & 1-3 & 66.1 & Non-function application & 1-3 & 17.2 \\
        {\tt then} and {\tt else}  & 1-3 & 73.1 & Difficult to explain  & 1-3 &
        25.8 &  Difficult to explain & 1-3 &  0.6\\
        Difficult to explain  & 1-3 & 20.9 & & & & & &\\
        \bottomrule
      \end{tabular}
    }
      \caption{Preliminary analysis of 2014 course}
      \label{tab:2014}
  \end{center}
\end{table}

\section{Evaluation} 

We evaluated how much better the extended type debugger is in
comparison with the previous one.
First we simulated what would have happened if we used the extended
type debugger for the 2012 course.
We then did a preliminary analysis of the extended type
debugger in the 2014 ``Functional Programming'' course.
We examined the difficult cases that we encountered in Section 7.1.

\subsection{Simulation of 2012 course}
Assuming that students answer the same as in the logs, we manually
simulated what would have happened if we had the extended type debugger for the
2012 logs (Table \ref{tab:simulation}).
The columns Conditional and Match in Table \ref{tab:simulation} summarize the results for the cases where the detected expressions are
ill-typed conditional and match expressions, respectively.
In both cases, we observe that the extended type debugger can provide
more specific error messages most of the time.
We describe the ``Difficult to explain'' cases in detail in Section 7.1.

For the cases where the detected expressions are ill-typed
applications, we observed that in 98.4\% of all cases,
the type debugger could show enough information from which students
should be able to understand the source of the type error.
The remaining cases, 1.6\% of all cases, are classified as difficult
(see Section 7.1).

\subsection{Preliminary analysis of 2014 course}
We used the extended type debugger in the 2014 ``Functional
  Programming'' course.
This year, {\em all} the user interactions, regardless of whether type
errors occur, are logged, so that we can analyze students behavior
more accurately.
To ease the classification efforts, we numbered the error messages to
indicate what the source of the type error is.
Table \ref{tab:2014} presents the initial results, showing
the number of times the classified errors occurred.

Most of the type errors in conditional expressions come from
type conflicts between \texttt{then} and \texttt{else} branches.
There are only 1.5\% of missing-\texttt{else} type error at language level
1.
Similarly, the sources of type errors for match expressions are mostly
from conflicting expressions.
In both cases, the type debugger can provide specific error messages
(except for ``Difficult to explain'' cases which are explained in the
next section).

For applications, ``Partial application'' means that some of
the arguments are higher order.
Since partial applications are not used at level 1, this means that
students did not pass enough arguments.
The table shows that students often pass fewer arguments than
necessary.
The ``Argument'' means that all the arguments are first order.
In this case, the type debugger displays all the type information of
the function and its arguments.

Interestingly, not a few type application errors come from
application of non-functions.
Typically, this happens when students forget ``;'' in a list
(e.g., \texttt{[2;\ 1;\ 4 5]} instead of \texttt{[2;\ 1;\ 4;\ 5]})
or when students forget an infix operator
(e.g., \texttt{"hello" "world"} instead of \texttt{"hello" \UP\ "world"}).
These cases are handled by the type debugger without any problems.
We separated these cases in order to give a slightly better error messages, saying
that the expression is parsed as an application but a function is missing.

Overall, compared with the original type debugger,
the extended version detects smaller expressions as
sources of type errors and gives more precise and informative error
messages. We have found that more students are trying to use the type debugger
compared with two years ago.

\section{Limitations}
Overall, our type debugger detects and expresses where the sources of
the type errors are, but we nonetheless found that it has some limitations.

\subsection{Difficult cases}
As we classify the logs, we encounter difficult cases where all the
subexpressions are well-typed and satisfy the necessary type constraints,
but the whole expression does not type check.
In all the cases, they have a variable with conflicting types.
Even though all the subexpressions are well-typed independently, the
same variable has one type in one subexpression and another type in
other subexpressions.
For example, consider the following expression: 
\begin{alltt}
fun p -> fun q -> if p and (q = 1) then p else 
\end{alltt}
After identifying the conditional expression as the source of the type
error, the extended type debugger checks whether subexpressions
satisfy the type constraints as follows: 
\begin{enumerate}
\item Annotate the predicate \texttt{p and (q = 1)} as \texttt{bool} and
pass it to the compiler's type inferencer.  It type checks with the
environment {\tt \{p :\ bool, q :\ int\}}.
\item Construct a list of branches \texttt{[p; q]} and pass it to the
compiler's type inferencer.  It type checks with the environment
{\tt \{p :\ 'a, q :\ 'a\}}.
\end{enumerate}
Thus, each subexpression of the conditional satisfies the required
type constraints independently.
However, the two environments conflict with each other and are not
unifiable.

It is not clear what informative error messages we can provide for such
cases.
Currently, our type debugger prints out that some variable is used at two
different types.
If we were to implement a dedicated type inferencer, we could say more
specifically what had happened in this case.
However, this is to the disadvantage of our type debugger, where the
compiler's type inferencer is reused, thereby making it easy to build a type
debugger that is consistent with the compiler's type inferencer.
It is not clear either if a dedicated type inferencer could actually
give a better error message.
In the future, we will investigate how serious this case is for novice
programmers and consider what we can do about it.

\subsection{Type variables}
In compositional type inference,
since a variable does not impose any constraints on its type, the type
of a variable always exists: a type variable.
Although the compositional type inference helps to identify
where the type of a variable is first forced to a specific type, showing
type variables in the error message can sometimes confuse students.
For example, the following program is taken from the 2012 logs:
\begin{alltt}
let rec search tree name = match tree with
        Empty -> Empty
      | Node (t1, st, n, t2) -> if (st = name) then n
                                else \fbox{search search t2
 name}\ensuremath{\sp{1}}
Error:
The first argument of this application causes a type
error. {\rm (highlight1)}
The types of the function, its arguments, and the required type
for the first argument are:
Function: 'a
First argument: 'b
Second argument: 'c
Third argument: 'd
Required for the first argument: 'e
\end{alltt}
The type debugger reports (without asking any questions) that the last
application is the source of the type error.
It does not ask any questions, because the application does not type
check but all its subexpressions trivially type check.
(Remember a variable always type checks.)
As a result, the extended type debugger lists all the arguments whose
types are all type variables.

One might think that we can at least say that the type of function and
the type of the first argument are the same type variable.
To do so, however, one is probably required to implement a dedicated
type inferencer.
Even if we could implement it, it is not at all clear how to explain
why the type error occurs.

\subsection{Users' input}
Our type debugger requires users' input.
This can be both an advantage and a disadvantage.
Without asking questions, it is impossible in general to locate a
single source of a type error that reflects the users' intentions.
By asking questions, our type debugger exactly locates the source of a
type error.
On the other hand, it means that the result depends on the users'
answer.
If a student inputs a wrong answer for some reason, e.g., as a result
of misunderstanding the questions, or a lack of knowledge on types,
the type debugger would identify a different place
as the source of the type error.
From our experience so far, we feel that questions raised by the type
debugger make students think about types of expressions more
seriously, which is nice from the pedagogic point of view.
However, we need more experience to draw a definite conclusion.

\section{Related work}

To produce understandable and appropriate error messages,
Heeren, Hage, and Swierstra \cite{Hage2003} divided type inference in the
Helium compiler for Haskell into two phases:
generation of type constraints and solving of type constraints.
When unsolvable type constraints are found during the constraint
satisfaction phase, they are output as type error messages.
By controlling which type constraint to remove from the unsolvable
ones,
we gain control over which expression to blame.
By registering a specific error message to each type constraint, we
can not only tailor type error messages \cite{Heeren2003}, but also
show possible fixes for typical error cases, such as proposing
similar `sibling' operators.
As such, Helium has been successfully used in a classroom setting with many
expression-specific error messages \cite{Hage2007}.

We share with Helium's authors the goal of producing better error messages.
To ascertain the programmers' intentions, we employ an interactive approach
where the debugger asks questions on the types of expressions.
Once the source of a type error is identified, the technique used
in Helium will be valuable to enhance error messages.

Interactive type debugging was proposed by Chitil \cite{Chitil2001}
for a small subset of Haskell. This was followed by Tsushima and Asai
\cite{Tsushima2012IFL} for full OCaml.
A similar approach was taken by Stuckey, Sulzmann, and Wazny
\cite{Stuckey2003}, who implemented an interactive type debugger for
Haskell (including various advanced features).
As in Helium, they produce type constraints and find the minimal
conflicting set to narrow down the possible cause of type errors.
By adding type constraints interactively, the programmer can express
his intention and be led to the sources of the type errors.

A conflicting set of type constraints is called a \emph{type error
slice} \cite{Haack2004}.
Among the type error slices, the minimal one is useful for type debugging
because it contains the smallest set of type constraints that lead to a type error.
The minimal type error slice can be automatically obtained without any
inputs on the programmer's intention.
Rahli, Wells, and Kamareddine implemented a type error slicer for
the full set of SML \cite{RWK2010}.

Writing a dedicated type error slicer is not an easy task.
Schilling \cite{Schilling2012} designed a type error slicer without
implementing a dedicated type inferencer but by reusing the compiler's
type inferencer.
Tsushima and Asai followed this approach \cite{Tsushima2013IFL}.
Lerner, Grossman, and Chambers used the compiler's type inferencer to
propose a possible fix to a type error.
They enumerated possible changes to an ill-typed program and check
whether the modified programs type check using compiler's type
inferencer as a black box.
If they type checked, they are shown as possible fixes of the type
error.

The compositional type inference used in our type debugger is based on
the one proposed by Chitil \cite{Chitil2001}.
Compositional type inference can identify where the type of a variable
is first fixed locally without being affected by the surrounding expressions.
A similar approach is used in the ${\cal U}_{AE}$ system by Yang \cite{Yang2000},
which can report which parts of a program have type conflicts.
Note that the ${\cal U}_{AE}$ system is in accord with most of the manifesto
for good error messages \cite{YMTW2000}.

There are also many studies and user tests about the reactions of
novice programmers while they are programming. Marceau et al.\
reported on novice programmers using Racket
\cite{Marceau2011.3,Marceau2011.10}.
They analyzed the effectiveness of each error message of the
Racket language.
We basically followed their advice in designing the error messages of our
type debugger. There is a report \cite{Laurie2008} that classified and discussed
the efficiency of debugging methods for Java programs (e.g.,
inserting printf, use of JavaDoc or debuggers, and narrowing erroneous
programs by commenting out).
While they focused on the methods of debugging, our user tests focused
more on the reactions of programmers. 
 Spohrer and Soloway analyzed bugs in
Pascal code written by novice programmers \cite{James1986}. They
classified bugs not only by expressions but also by the users' intention
(``plan''). They categorized bugs into ``correct plan but
wrong implementation'' and ``wrong plan and implementation''. This
paper also focused on the intentions of users in order to detect the
sources of type errors. 

The idea of pedagogical language levels is attributed to
Felleisen, Findler, Flatt, and Krishnamurthi \cite{Felleisen2004}.
They proposed an alternative
role for functional programming for novices.
The idea of dividing a language into levels is widely used in the
Racket language.
It has many useful language levels from beginners to experts.
The Racket language levels are used not only for providing better
error messages but also for avoiding unnecessary advanced syntax,
among others.
We imported this idea to improve the error messages of the type
debugger.
The Helium compiler employs simple language levels via a flag.
When the flag is off, overloading is turned off and error messages
no longer mention type classes.

\section{Conclusion and future work}

In this paper, we analyzed the logs of our interactive type debugger
from the 2012 functional language course in our university.
Although the type debugger worked well for some cases,
the analysis showed that expression-specific and more informative
error messages are desired.
According to these observations, we extended the type debugger to
provide expression-specific error messages for conditional and match
We evaluated the resulting type debugger with logs from 2012
and showed a preliminary analysis of the logs from the 2014 course.
In both cases, we showed that the type debugger provides
more informative error messages in most of the cases.
We also showed difficult cases where a variable was used in
conflicting types.

Although a thorough evaluation of the type debugger requires a full analysis of the logs from the 2014 course, we have
the impression that the type debugger deals with most of the common
type errors properly.

We plan to further pursue the following
directions in the future.
First, we want to analyze the new logs from the 2014 course
and see if there are cases the current type debugger
cannot handle well.
We already found some cases where expression-specific handling is
required, such as \texttt{try with} blocks.
Supporting them is essential to making the type debugger robust.
On a related note, an automatic log analysis could be an interesting
topic.
The amount of logs is becoming large and it is simply unrealistic to
manually analyze them.

Once the type debugger covers most of the common errors,
we want to build a taxonomy of common type errors
of novice programmers.
We could then create a document describing typical errors, which
novices can study.
In particular, we found type errors that can be identified by the type
debugger but more information in the error message would be useful.
For example, not a few students write only an exception (e.g.,
\texttt{Not\US found}) without raising it.
Another example is to forget writing a dereference (\texttt{!}) before a
reference.
In such cases, the type debugger could show typical related mistakes.

We also plan to use the type debugger in the upper level course to see
if it is useful for medium-level programmers.
In that case, the programs to be debugged are large and the number
of questions will increase.
To reduce the number of questions, we could introduce type-error
slices.
The type debugger is not only for novice programmers: it is also for
experts to use.

\section{Acknowledgement}
We appreciate the valuable feedback and constructive comments by the
reviewers.

\bibliographystyle{eptcsini} % alphabetical order of the family name of the author
\bibliography{generic}

\end{document}